\documentclass[11pt,a4paper]{article}
\usepackage{amsfonts}
\usepackage{amsmath,amsfonts,amsthm,mathrsfs}
\usepackage{graphicx}
\usepackage{verbatim}
\usepackage{color}
\usepackage{picinpar}
\usepackage{booktabs}
\usepackage[doublespacing]{setspace}
\usepackage[authoryear]{natbib}
\usepackage{tabularx}

\makeatletter
\newcommand{\figcaption}{\def\@captype{figure}\caption}
\newcommand{\tabcaption}{\def\@captype{table}\caption}
\makeatother

\theoremstyle{definition}   %

\theoremstyle{remark}

\numberwithin{equation}{section}

\def\eop{{\hfill\vbox{\hrule height .3pt
      \hbox{\vrule width.3pt height 7pt
      \kern 7pt
      \vrule width .3pt}
      \hrule height .3pt}} \par\bigskip}

\title{ Financial Market Directional Forecasting With Stacked Denoising Autoencoder }
\author{ Shaogao Lv$^{a}$,Yongchao Hou$^{b, *}$, Hongwei Zhou$^{b}$,
\\
{\small $^{a}$ Department of Statistics and Mathemtics, Nanjing Audit University, Nanjing,  China;} \\
{\small $^{b}$ Statistics School, Southwestern University of
	Finance and Economics, ChengDu,  China;} \\
{$^{*}$} Corresponding author. Email address: }
\date{}
\begin{document}
\maketitle
\doublespacing

\setcounter{page}{1}
\begin{abstract}
  Forecasting stock market direction is  always an amazing but challenging  problem in finance. Although many popular shallow computational methods (such as Backpropagation Network and Support Vector Machine) have extensively been proposed, most algorithms have not yet attained a desirable level of applicability.
   In this paper, we present a deep learning model with strong ability to generate high level
   feature representations for accurate financial prediction. Precisely,
   a stacked denoising autoencoder (SDAE) from deep learning is applied to predict  the daily CSI 300 index, from  Shanghai and Shenzhen Stock Exchanges in China. We use six evaluation criteria to evaluate its performance compared with the back propagation network, support vector machine. The experiment shows that the underlying financial model with deep machine technology has a significant  advantage for the prediction of the CSI 300 index.
\end{abstract}

{\bf Key Words and Phrases:}  Deep Learning, Stacked Denoising Autoencoder, Neural Networks, Support Vector Machine(SVM), Financial Market Forecasting.

\bigskip

\section{Introduction}
Forecasting the financial market is full of big challenge in both academia and business. Because
of the noisy nature and dynamic mechanism, it is quite difficult to forecast the true signals from financial time series. This naturally leads to the debate on market predictability among the academics and
market practitioners. It is known that the efficient market hypothesis \citep*{Fama1970} and the random
walk hypothesis \citep*{Malkiel1973} are major theories in economics and finance. The hypotheses
state that the financial market evolves randomly and no excess returns can be earned by
predicting and timing the market. According to these hypotheses, any technique analysis cannot  consistently
outperform the market, and the simple ``buy-and-hold" is the best investment strategy.
However, many investors and scholars are opposed to these hypotheses, and they believe that the financial market is predictable to some extent. There has been a lot of related work disputing these hypotheses in the last decade, and particularly \cite{Hirshleifer2001}
provided a survey of empirical evidence for the capital market inefficiency, as well as reviewed the
explanation for these findings from the  perspective of behavioral finance.

During the past decade, researchers in the machine learning and data mining community have also tried to forecast the financial market, using various learning algorithms.
 Artificial neural networks (ANN) have been successfully used
for modeling financial time series \citep*{Choi1995,Zhang1998,Kaastra1995,Chiang1996}.
 Unlike traditional statistical models, neural networks, as
 a class of data-driven and nonparametric weak models, can approximate  any nonlinear
function without a priori assumptions about data structure.  As a result, ANN are insensitive to the problem of model misspecification unlike many ordinary statistical methods. Besides the above work, some researchers  are more interested in  combining additional techniques with ANN, to further improve the prediction ability. For example, \cite{Tsaih1998} integrated the rule-based
technique and ANN to predict the direction of the S\&P 500 stock index futures on a
daily basis. \cite{Kim2000} integrated genetic algorithm with ANN to predict the stock price index.
They first proposed a genetic algorithm approach to discrete features,  and then used ANN to determine the connection weights. However, some of previous studies show that ANN had some  poor performance at learning
the underlying patterns, probably because stock market data has tremendous noise and complex rules.
 Moreover, Back Propagation Neural Network(BPNN) often suffers a lot from selecting a large number of tuning parameters,
including  hidden layer size, learning rate and momentum term.

On the other hand, a novel type of learning machine, called  support
vector machine (SVM), has been receiving increasing attention in various areas including quantitative finance.
SVM was developed by Vapnik and his colleagues \citep*{Vapnik1995}. Essentially, many traditional neural network models conform to  {\it empirical risk minimization} principle, whereas SVM implements the {\it structural
risk minimization} principle. In contrast to ANN estimation, the SVM solution can be derived from convex optimization,  making the optimal solution both global and unique.

There has been a few studies that deal with stock market directional forecasting using
 SVM classification techniques. \cite{Kim2003} showed that the SVM outperformed the ANNs in predicting future
direction of a stock market and yet reported that  SVM can attain the  prediction accuracy of 57.8\% in the experiment with the Korean composite stock price index 200. \cite{Huang2005}  used a SVM-based model to predict Nihon Keizai Shimbun Index 225 in a single period, and reported that the prediction accuracy of 75\%  can be achieved. \cite{Wang2013} proposed a dimension-reduced SVM  via PCA to predict the upward or downward direction of Korean composite stock price index  and Hangseng index. Their experimental results show notably high hit ratios in predicting the movements of the individual constituents. More importantly, \cite{Wang2010} proposed a new SVM classifier with multiple kernels and illustrated sufficiently the importance of kernel selections to stock index forecasting. However, 
a major drawback of SVM for the direction prediction is that the input variables lie in a high-dimensional feature space, ranging from hundreds to thousands. Moreover, the storage of large matrices requires a lot of memory and computational cost. As ANN, SVM also belongs to shallow learning approaches, with limited learning ability and suffers  a lot from noise interference in financial time series.

Deep learning,  a new area of machine learning,  aims at building deep architectures so as to   
represent well  the characteristics within data. For the learned representation, the lower-level features
represent basic elements or edges in smaller area of data, while the higher-level features
represent the abstract aspects of information among data. Theoretical results have suggested that deep learning architectures with multiple levels of non-linear operations provide high-level abstractions for object recognition similar to those found in the human brain. 

Various deep learning architectures have been proposed recently, such as deep belief networks \citep*{Hinton2006}, convolutional deep neural networks \citep*{Lee2009}, denosing autoencoders \citep*{Vincent2008}. See a related survey of deep learning \citep* {Bengio2009}. There have been  a lot of successful applications in computer vision, automatic signal recognition, and natural language processing, as well as climate forecasting \citep*{Chen2012}. In the last few years, there has also been serval applications of deep learning implemented  in the area of finance,  such as \citep*{Ribeiro2009,Takeuchi2013, Yeh2014}.

This paper attempts to provide a new perspective on the financial market prediction problem using
deep learning algorithms. In view of nonlinearity highly and tremendous noise among data,  we make full use of the advantages of SDAE to overcome them to some degree. We consider  the upward or downward direction on a collection of daily returns of the  CSI 300 index from 2005 to 2014.  In our experiments, for comparison,
we treat the results of models training via the traditional BPNN and SVM classifier as baselines. The results
shows that the deep learning algorithm (SDAE) significantly outperforms the two baselines. In addition to
the superior performance, more importantly, the representation of data can be automatically
generated during the learning process.

The rest of the paper is organized as follows. In Section 2, BPNN, the standard SVM and the deep learning algorithm for SEAD are briefly described. The Section 3 presents experimental setup,
 which examines the CSI 300 index and gives the
evaluation metrics for problem assessment. In Section 4 we compare the performance of different methods based
on historical data of the CSI 300 index. Finally  we conclude the paper in Section 5.

\section{Analytical Methods}
A branch of statistical learning (or machine learning) is mainly concerned with the development of proper refinements of the
regularization and model selection methods in order to improve the predictive
ability of algorithms. This ability is often referred to as generalization,
since the algorithms are allowed to generalize from the observed training data to new data.
 One crucial element of the evaluation of the generalization ability of a particular model is the measurement of
the predictive performance results on out-of-sample data, i.e., using a collection of data, disjoint from the in-sample data that has already
been used for model parameter estimation. We provide some brief descriptions of three methods in this section, and focus more
on one of deep learning algorithms, the SDAE adopted in this paper.

\subsection{Backpropagation Neural Networks}
The BPNN is a kind of multilayer feed-forward networks with training by error
backpropagation algorithm \citep*{Zhang2002}. It is a family of supervised learning, and its
idea behind  it applies the  gradient descent method to reach a given
accuracy approximation to some unknown function. The classical BP neural network consists of a three-layer structure: input-layer nodes, hidden-layer nodes and output-layer nodes. Usually, BP neural networks are fully connected, layered,
feed-forward models, and the so-called {\it activations} flow from the input layer through the hidden layer, then to the output layer. To derive appropriate weights among networks,
the BP network often begins with a random set of weights, and then the network adjusts its current weights at each iteration using all input-output pairs. Each pair is dealt with at two stages, a forward pass and a backward pass respectively. The forward pass involves presenting a sample input to the network and letting activations flow to the output layer. During the backward pass, the  actual output in network is compared with the target output and error estimates are computed for the output units. The weights connected to the output units are adjusted accordingly by a gradient descent method. The error estimates of the output units are
then used to derive error estimates for the units in the hidden layer. Finally, errors are propagated back to the nodes stemming from the input units. Repeating this process, the BP network updates its weights incrementally until the network converges. Although those neural networks can capture nonlinear efficiently, the main drawbacks for neural networks are that only local solutions are found, and also tend to lead to overfitting. Due to these limitations, learning capacity even for multiple layer networks cannot be improved significantly. For further details about classical neural networks, we refer the readers  to reference \citep*{Bishop1995}.

\subsection{Support Vector Machine}
Support vector machines  are a class of popular learning machines with  shallow architectures, proposed originally by  \cite{Vapnik1995}. They are based on the {\it structural risk minimization} principle from the
perspective of statistical learning theory. More precisely,  SVM  corresponds to a specific linear method in a high dimensional feature space, but it is highly nonlinear with respect to the original input space.
Essentially,
SVMs are often regarded as extensions of a large class of neural nets, radial basis function  nets, and polynomial classifiers. It is known that theoretical foundations for SVM have been established mathematically,
as well as their numerical procedures are often computational efficiently, and robust to the dimension of input space.

 Now let us recall the standard SVM classification on binary data. Assume that there is an input space, denoted by $X\subseteq \mathbb{R}^d$,  and  a corresponding  output
space denoted by $Y=\{-1,1\}$. Given the training sample  $\mathrm{D}$ is available from some underlying distribution, where $\mathrm{D}=\big((x_1, y_1), (x_2, y_2),....,(x_n, y_n)\big)$ and $n$ is the sample size.
the task of classification is to construct a heuristic
function $f(x)$, such that $sign(f(x))\approx y$ over the whole population distribution.
To handle nonlinear problems, a nonlinear map $\phi:X\rightarrow \mathcal{F}$ is introduced in advance, where $\mathcal{F}$  is called the feature space (with high dimensions). Then we attempt to find such a classifier that maximizes  the  margin distance over $\mathcal{F}$. Equivalently, SVM classification problem can be reduced to the following constrained convex programme,
\begin{eqnarray}\label{equ}
 && \min_{w,b,\xi}\langle w,w\rangle+C\sum_{i=1}^n\xi_i\\
 &&\hbox{subject to}\quad y_i(\langle w,\phi(x_i)+b\rangle)\geq 1-\xi_i,\,\xi_i\geq 0,i=1,2,...,n,
\end{eqnarray}
where $C$ is a penalty parameter trading off variance and bias.
By the dual transformation,  we can rewrite the above problem as the following form
 \begin{eqnarray}
&&\max_{\alpha} \Big\{\sum_{i=1}^n\alpha_i-\frac{1}{2}\sum_{i,j=1}^ny_iy_jK(x_i,x_j)\Big\}\\
&&\hbox{s.t.}\quad \sum_{i=1}^ny_i\alpha_i=0,\,\,0\leq \alpha_i\leq C,\,i=1,2,...,n,
\end{eqnarray}
where $K(x,u)=\langle\phi(x),\phi(u)\rangle$ is called a kernel function, and Guassian kernels
($K_\sigma(x,u)=\exp(-\|x-u\|^2/\sigma^2)$) have been used widely for various problems.
Note that this dual programme belongs to  quadratic convex optimization, and many existing approaches including interior point methods can solve it efficiently. For the detailed contents for SVM, please refer to a representative book related to SVM \citep*{Scholkopf2002}.

\subsection{Stacked Denoising Autoencoders}
Theoretical results suggest that deep learning architectures with multiple levels
of non-linear operations provide high-level abstractions for object recognition similar to those found in the human brain. Until now, there has been various popular deep learning algorithms developed in recent years, such as deep neural networks, convolutional  neural networks, and restricted Boltzmann machines.
Moreover, several deep learning algorithms have been applied to fields like computer vision, automatic speech recognition, natural language processing and audio recognition, and they have been shown to be comparable performance than  state-of-the-art results on various tasks.

The SDAE  is an extension of the stacked autoencoder \citep*{Bengio2007} and it was introduced in \cite{Vincent2008}. As stated briefly in \cite{Bengio2009}, the traning procedure for SDAE is mainly listed as follows:

1. Train the first layer as an autoassociator to minimize some form of reconstruction error of the raw
input. This is purely an  unsupervised learning.

2. Let the hidden units' outputs in the autoassociator, derived as above,  be an input for another layer, and thus another autoassociator is generated naturally. Note that, we only need unlabeled examples.

3. Iterate as in (2) to add the desired number of layers.

4. Take the last hidden layer output as input to a supervised layer and initialize its parameters.

5. Fine-tune all the parameters of this deep architecture  based on the supervised criterion. If possible, unfold all the autoassociators into a very deep autoassociator and fine-tune the global reconstruction
error.

From the above steps, we can see that the SDAE consists of two key facades: a list of autoencoders, and a multilayer perceptron. During pre-training we use the first facade, that is, we treat our model as a list of autoencoders, and train each autoencoder separately. In the second stage of training, we use the second facade. Essentially, these two facades are linked together, since the autoencoders and the sigmoid layers of the MLP share parameters. In practice, this greedy layer-wise procedure has been shown to yield significantly better
local minima than random initialization of deep networks, achieving better generalization on a number of tasks.

\section{ Experimental Setup}
In this section, we apply the stacked denoising autoencoder to forecast the financial market
and compare its performance to two other methods, including BP network and SVM.

\subsection{Dataset}
The experiments are based on  technical indicators, the market price and the direction of change in the daily CSI 300 index. The CSI 300 is a capitalization-weighted stock market index, designed to replicate the performance of 300 stocks traded in  Shanghai and Shenzhen Stock Exchanges.
The direction of daily price change of CSI 300 index which we attempt to forecast is used as output variables, and technical indicators and the market price in history are used as input variables. Note that technical indicators currently popular in stock market can be classified into five categories: moving averages, trend detection, oscillators, volume and momentum, and our initial features consist of  28 features selected from the above five categories, as well as open price, close price, high price, low price and volume. Table 1 shows all the variables used in our experiment.

\begin{center}
\tabcaption{ All the variables used in the experiment}
\begin{tabular}{cccccccc}

\toprule[1pt]
Categories      & Name of each attribute \\
\midrule[1pt]
Moving averages	     & EMA, SMA, EVWMA, ZLEMA, TRIX, MACD	 \\

Trend detection     & EMV, DEMA, ADX, AROON, CCI, TDI, VHF, DPO, ZigZag	    \\

Oscillators        & RSI, ATR, Volatility, ROC, CMO, MFI, WPR    \\

Volume        & OBV, CLV, CMF, ChaikinAD	   \\

Momentum    & Momentum, Stoch   \\

Price Volume  & Open, Close, High, Low, Volume	   \\
\bottomrule[1pt]
\end{tabular}
\end{center}
		
Since the main goal in this paper is to predict the directions of daily change of the  CSI 300 index, they belong to be binary classification problem,  and we denote by ``-1" whenever the next day's index is lower than today's index, and similarly we denote by ``+1" whenever the next day's index is higher than today's index.
The experiments were based on historical prices of the  CSI 300. All these methods are conducted on 2,289 trading days from  August 2, 2005 to December 31, 2014, which covers for around 9 years.
Considering that the stock market is a time-varying market, we use the sliding window to make rolling forecast. Moreover, choosing an appropriate forecasting horizon is quite critical in
financial forecasting. From the trading aspect, the forecasting horizon should be sufficiently long so that the
common underlying pattern may exist over different periods. From the prediction aspect, the forecasting horizon should be short enough due to  the limited persistence of financial time series.
 To this end, we first select the first 1400 trading days as the training sample, and the next 100 trading days as test samples. Then we remove the first 100 days from the above 1400 trading days, so that the remaining 1300 data and the above 100 test sample data consist of the new training sample, and repeated the same procedure. In this way we finally can obtain 9 time intervals for predicted results. In our procedure,
the original numerical data are scaled into the range of $[0, 1]$, so as to ensure that the input attributes with large values do not overwhelm the attributes with small values.

\subsection{ Evaluation Criteria}
In order to make a more reasonable assessment of the quantitative timing strategy, this paper will evaluate the experimental results and the investment performance for these three models.
Firstly, like traditional classification tasks, we evaluate the experimental results with {\it accuracy, precision, recall and F-score}. These four quantities  are utilized to measure the performance of positive and negative class respectively, derived from a confusion matrix that records corrected and uncorrected examples for each class, see details as follows.
\begin{center}
\tabcaption{ A confusion matrix for binary classification}
\begin{tabular}{cccccccc}

\toprule[1pt]
     &\hspace*{-1cm}Positive\,\,\,\hspace*{1cm}	Negative \\
\midrule[1pt]
True	     & True Positive(TP), 	True Negative(TN)	 \\

False     & False Positive(FP),	False Negative(FN)	.   \\

\bottomrule[1pt]
\end{tabular}
\end{center}

Table 2 presents a confusion matrix for binary classification, where TP, FP, FN and TN   represent {\it true positive,  false positive,  false negative and  true negative}, respectively. {\it Accuracy} assesses the overall effectiveness of some given model. It is given by
$$
\hbox{{\it Accuracy}=(TP+TN)/(TP+FP+FN+TN)}.
$$
{\it Precision} evaluates the right information given by the model, expressed by
$$
\hbox{{\it Precision}=TP/(TP+FP)}.
$$
{\it Recall} can be understood as the ratio of the correct positive class information and the actual class information given by the model, expressed by
$$
\hbox{{\it Recall}=TP/(TP+FN)}.
$$
{\it F-score} is the product of Precision and Recall,  given by
 $$
\hbox{{\it F-score=Precision$\times$ Recall}}.
$$
Next, this paper uses the {\it transaction success rate, cumulative return, maximum drawdown} to evaluate  investment performance of trading strategy. {\it Transaction success rate} is the ratio of the number of the positive return divide by the total number of transactions, after deducting the fee and the impact cost induced by the model. It is given by
$$P=\frac{t}{T},$$
where $T$ is the total number of transactions, and $t$ is the number of the positive return after deducting the fee and the impact cost.
		
Presented as a percentage, the {\it cumulative return} of some security is the raw mathematical return of the following calculation:
$$
cumulative \,\,return=\frac{current\,\, price-orginal \,\,price}{orginal\,\, price}.
$$

The {\it  maximum drawdown}, as a risk metric,  measures the peak-to-trough loss of an investment.
It offers investors a worst case scenario, and tells the investor how much would have been lost if an investor bought at the absolute peak value of an investment. More formally,
$$
MMD(T)=\max_{\tau\in (0,T)}\Big\{\max_{t\in (0,T)}X(t)-X(\tau)\Big\},
$$
where $X(\cdot)$ typically represents the cumulative return of some security.

\section{ Experimental results and analysis}

In this subsection, the purpose of the following experiment is to compare
SDAE with the standard SVM, as well as the BP neural
network with there layers.
In our experiment, the Gaussian radial basis function are used as the kernel function of  SVM. For the tuning parameter $\sigma^2$ in the Guassian kernel, we first computed the median of $\|x_i-\bar{x}\|_2$, where $x_i$
is the input feature vector of the $i$-th training observation and $\bar{x}$ is the corresponding mean.
Denote it as $m_\mathbf{x}$, and then we search the optimal $\sigma^2$ over $\{0.2m_\mathbf{x}, 0.4m_\mathbf{x}, 0.6m_\mathbf{x},...,2m_\mathbf{x}\}$. In addition, the regularization parameter $C$ in SVM is determined by the standard cross validation.

 A standard three-layer BP neural network is used as a benchmark. There are 28 nodes in the input layer, which is equal to the number of indicators. The number of hidden nodes is determined based on the validation
set, which are both 50 in this experiment. The BP software used is directly taken from Matlab  neural network toolbox.

Tables 3-6 show the predicted results on test data of these three models. From these tables, in terms of all the above five evaluation criteria, we find that SDAE has a comparable performance to these of BPNN and SVM.
In particular, The SDAE achieves an overall accuracy rate $65.5\%$, in contrast, two accuracies $51.4\%$ and $61.2\%$ are achieved by BPNN and SVM respectively. The result means that SDAE
can forecast more closely to the actual values of  index change direction than two other standard methods.
Note that, it is seen from Table 4-6, BPNN has a very poor performance for the actual positive sample.
The same conclusions also hold in terms of the other four criteria and see the corresponding tables for the details. That being said, the SDAE that can realize more complexity than the shallow layer methods  (i.e.,
BPNN and SVM) have had even better performance, and demonstrate the power of deep learning.

On the other hand, Table 7, 8 and Figure 1 just show the investment performance of SDAE and SVM, since BPNN has a poor performance compared with  SDAE and SVM in a whole. From table 7, we can see that the transaction success rate of SDAE is 50.6\%, which is better than that of  SVM (45.7\%). Furthermore, the SDAE has an absolute advantage in term of  the cumulative return in compared with SVM, which is also indicated in Figure 1. Actually, this is also a main concern for many  investors.
In addition, Table 8 lists the biggest five drawdown induced by SDAE and SVM, from which we find that the depth and length of drawdown of SDAE are better than SVM. More precisely, the biggest drawdown of SDAE is 18.92.
In other words, there is a smaller probability in which an extreme loss happens, by contrast  with two other methods.

\begin{center}
\tabcaption{ The prediction accuracy of BPNN, SVM and SDAE}
\begin{tabular}{cccccccc}

\toprule[0.5pt]
   Period  & BPNN(\%)	&SVM(\%)	&SDAE(\%)	 \\
\midrule[0.5pt]
1	     & 54& 	64& 67	 \\

2     & 48	& 57 &60  \\
3	&55.0	&66.0	&69.0\\
4	&48.0	&57.0	&67.0\\	
5	&46.0	&61.0	&66.0	\\
6	&56.0	&55.0	&59.0	\\
7	&44.0	&65.0	&63.0	\\
8	&50.0	&58.0	&65.0	\\
9	&62.9	&68.5	&74.2	\\
Average	&51.4	&61.2	&65.5\\
\bottomrule[0.5pt]
\end{tabular}
\end{center}

\begin{center}
\tabcaption{ The prediction precision of BPNN, SVM and SDAE}
\begin{tabular}{cccccccc}

\toprule[0.5pt]
   Period  & BPNN(\%)	&SVM(\%)	&SDAE(\%)	 \\
\midrule[0.5pt]
1&	0&	61.90&	60.66\\	
2&	48.0&	53.73&	59.09\\	
3&	0&	64.86&	69.44\\	
4&	48.0&	55.10&	71.43\\	
5&	0&	68.29&	71.74\\	
6&	51.72&	50.00&	53.03\\	
7&	44.0&	60.00&	55.38\\	
8&	0&	61.11&	63.64\\	
9&	62.92&	71.21&	76.19\\	
Average& 50.48&	60.22&	63.69\\	
\bottomrule[0.5pt]
\end{tabular}
\end{center}

\begin{center}
\tabcaption{ The prediction recall of BPNN, SVM and SDAE}
\begin{tabular}{cccccccc}

\toprule[0.5pt]
   Period  & BPNN(\%)	&SVM(\%)	&SDAE(\%)	 \\
\midrule[0.5pt]
1&	0&	56.52&	80&43\\	
2&	100&	75.00&	54.17	\\
3&	0&	53.33&	55.56\\	
4&	100&	56.25&	52.08\\	
5&	0&	51.85&	61.11\\	
6&	33.33&	68.89&	77.78\\	
7&	100&	61.36&	81.82\\	
8&	0&	44.00&	70.00\\	
9&	100&	83.93&	85.71\\	
Average&	48.39&	61.47&	68.81\\	
\bottomrule[0.5pt]
\end{tabular}
\end{center}

\begin{center}
\tabcaption{ The prediction F-score of BPNN,SVM and SDAE}
\begin{tabular}{cccccccc}

\toprule[0.5pt]
   Period  & BPNN(\%)	&SVM(\%)	&SDAE(\%)	 \\
\midrule[0.5pt]
1&	0&	34.99&	48.78\\	
2&	48.00&	40.30&	32.01\\	
3&	0&	34.59&	38.58\\	
4&	48.00&	30.99&	37.20\\	
5&	0&	35.41&	43.84\\	
6&	17.24&	34.44&	41.25\\	
7&	44.00&	36.82&	45.31\\	
8&	0&	26.89&	44.55\\	
9&	62.92&	59.77&	65.31\\	
Average&	24.43&	37.02&	43.83\\	
\bottomrule[0.5pt]
\end{tabular}
\end{center}

\begin{figure}
  \centering
   \includegraphics{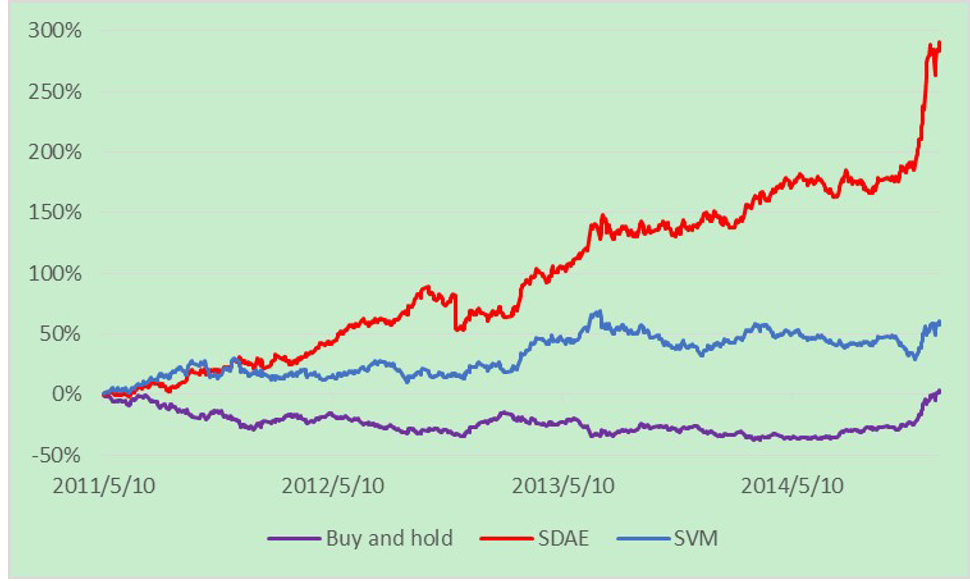}
     \caption{The cumulative return of SDAE and SVM.}
    \label{fige2}
\end{figure}

\begin{center}
\tabcaption{ The transaction success rate and cumulative return of SDAE and SVM}
\begin{tabular}{cccccccc}

\toprule[0.5pt]
    & SDAE	&  SVM \\
\midrule[0.5pt]
Transaction success rate(\%)&	50.6&	45.7	\\
Cumulative return(\%)&	291&	61	\\
\bottomrule[0.5pt]
\end{tabular}
\end{center}

\begin{center}
\tabcaption{ The biggest five drawdown of SDAE and SVM}
\begin{tabular}{cccccccc}

\toprule[0.5pt]
  	Stage&   Begin  &	Bottom &	 End  &	      Depth &	Length &	Fall   &	Recovery\\	
           1(SDAE) &	  2012-10-09 &	2012-11-23 &	2013-03-07 &	-18.92 &	100 &	34 &	66	\\
	          2(SDAE) &	2013-07-15 &	2013-07-29 &	2013-12-20 &	-8.22 &	108 &	11 &	97	\\
	        3(SDAE) &	2011-08-08 &	2011-08-22 &	2011-09-19 &	-8.06 &	30 &	    11 &	19\\	
	          4(SDAE)&	2011-12-14 &	2012-01-05 &	2012-02-08 &	-7.45 &	34 &	    15 &	19\\	
	         5(SDAE) &	2014-05-26 &	2014-07-16 &	2014-08-04 &	-6.74 &	50 &  	37 &	13	\\
             1(SVM) &	 2013-07-09 & 2014-11-21 &	2011-11-21 &	-23.79 &	365 &   336 &	332	\\
 	         2(SVM) &	2011-12-06 &	2012-09-05 &	2011-12-02 &	-15.06 &	299 &	184 & 115 \\
 	         3(SVM) &	2011-10-19 &	2011-11-09 &	2011-12-02 &	-11.60 &	33 &	    16 &	17	\\
 	         4(SVM) &	2013-04-24 &	2013-05-14 &	2013-06-07 &	-5.33 &	30 &	    12 &	18	\\
 	        5(SVM) &	2011-09-09 &	2011-09-19 &	2011-09-22 &	-5.17 &	9 &	    6 &	3	\\
\bottomrule[0.5pt]
\end{tabular}
\end{center}

\section{Conclusion}
We present the deep learning model (Stacked Autoencoder) for the direction prediction of the  CSI 300 index  and discuss the results comparing two different methodologies: BBNN and SVM. The preliminary results with a deep learning architecture are promising and raise interest regarding its application to this problem. The trading strategy based on our method consistently outperforms the market, and the excess returns are statistically significant. Essentially,
the properties of the SAED model allow extracting a high-representation of the features, that may describe the
financial status of stock market through a greedy layer-wise unsupervised learning. Meanwhile, by means of dimensional reduction of  the SAED, a large amounts of  noise in financial data can be removed, so that it is likely to find the underlying common pattern  among stock data. Finally, this demonstrates that the financial
market is not completely efficient and is predictable to some extent, which is in accord with
the conclusions made by many previous work.

\bigskip
\bigskip

{\bf Acknowledgement.}
The  second author's research is supported
partially by National Natural Science Foundation of China (Grant No.11301421),
 and Fundamental Research Funds  for the Central Universities of China (Grants No.
JBK141111, 14TD0046 and JBK140210).


\end{document}